\documentclass[twocolumn,preprintnumbers,elsart]{revtex4}
\usepackage{eurosym}
\usepackage{makeidx}
\usepackage{amssymb}
\usepackage{amsmath}
\usepackage{mathrsfs}
\usepackage{graphicx}
\usepackage{dcolumn}
\usepackage{bm}
\usepackage[center]{subfigure}
\usepackage{color}
\usepackage{indentfirst}
\usepackage{array}
\usepackage{booktabs}
\usepackage{multirow}
\usepackage{upgreek}
\usepackage[section]{placeins}

\begin{document}

\title{Elongated vortex quantum droplets in binary Bose-Einstein condensates}
\author{Guilong Li$^{1}$}
\thanks{These authors contributed equally to this work.}
\author{Zibin Zhao$^{1}$}
\thanks{These authors contributed equally to this work.}
\author{Rui Zhang$^{1}$}
\author{Zhaopin Chen$^{2}$}
\author{Bin Liu$^{1}$}
\email{binliu@fosu.edu.cn}
\author{Boris A. Malomed$^{3}$}
\author{Yongyao Li$^{1,4}$}
\email{yongyaoli@gmail.com}
\affiliation{$^{1}$School of Physics and Optoelectronic Engineering, Foshan University,
Foshan 528225, China\\
$^{2}$Physics Department and Solid-State Institute, Technion, Haifa 32000,
Israel\\
$^{3}$Instituto de Alta Investigaci\'{o}n, Universidad de Tarapac\'{a},
Casilla 7D, Arica, Chile\\
$^{4}$Guangdong-Hong Kong-Macao Joint Laboratory for Intelligent Micro-Nano
Optoelectronic Technology, Foshan University, Foshan 528225, China}

\begin{abstract}
Stability of elongated (``slender") quantum droplets (QDs) with embedded
unitary and multiple vorticity is a problem that was not solved previously.
In this work, we propose a solution which relies upon the use of the spatial
modulation of the inter-species scattering length in the binary
Bose-Einstein condensates, in the form of a two-dimensional axisymmetric
Gaussian, shaped by means of the optical Feshbach resonance. The
corresponding effective nonlinear trapping potential supports completely
stable elongated QDs with vorticity $S=0$ and partly stable families of
elongated QDs with $S=1,2,3,4$ (other nonlinear systems do not maintain
stability of vortex droplets with $\geq 2$). We systematically analyze
effects of the amplitude and width of the Gaussian modulation, as well as
the total number of atoms, on the shape and stability of the QDs, some
effects being explained analytically. Collisions between identical QDs with $%
S=1$ moving in opposite directions along the central axis leads to their
merger into still more elongated breathing QDs with the same vorticity,
while collisions between QDs with $S=\pm 1$ are quasi-elastic. Moving
modulation profiles are able to adiabatically rotate the trapped elongated
QDs. Application of a torque to the vector QD sets in the gyroscopic regime
of robust precession, which realizes a macroscopic spin-orbit-coupling
effect.
\end{abstract}

\maketitle
\section{Introduction}
Vortices with an elongated
three-dimensional (3D) structure have drawn much interest in astrophysics
\cite{McWilliams1994a,astro} and geophysics \cite{McWilliams1994b}, plasmas
\cite{plasma1,plasma2,plasma3}, superconductors \cite%
{Dolan1989,Odobesko2020,Hung2012,Olszewski2018}, optics \cite%
{Freund1997,Sharma2011}, and Bose-Einstein condensates (BECs) \cite%
{Stockhofe2011,Liu2013,McEndoo2010,OKtel2004,Engels2002,Tikhonenkov2008,Tan2023}%
. However, the realization of elongated self-localized
vortices - those stabilized solely by intrinsic interactions - has remained
elusive. Quantum droplets (QDs)
\cite{LHY,Petrov2015,Petrov2016,Luca1,Jrgensen2018,Cabrera2018,Ferrier-Barbut2018,Cheiney2018,Inguscio2018,Luca2,Yin2021,dlw2021,Luo2021,Cui2025} offer a promising platform to address this challenge, as they inherently support the formation of self-localized vortices without external linear confinement, stable 2D and 3D QDs with embedded vorticity and more complex vortex patterns have been predicted \cite{Kartashov2018,QDvortex2D,Tengstrand2019,Chen2024,Li2024FOPview}.

The most intuitive solution lies in dipolar QDs where the
stabilization emerges from the competitive balance between anisotropic
dipole-dipole interactions (DDIs) \cite{Stuhler2005,Lahaye2009} and beyond
mean-field effects. Although such systems exhibit intrinsic anisotropy
\cite{Ferrier-Barbut2016,Schmitt2016,Santos,Wang2020,Bottcher2021,Yin2022},
elongated vortex structures have proven fundamentally unstable
\cite{Cidrim2018}, permitting only short-axis-aligned vortex configurations
which the vorticity axis is perpendicular to the direction of the dipolar
polarization \cite{Li2024PRL,Li2024FOP}.

An alternative approach within quantum droplets emerges from two-component atomic systems, which can sustain stable vortices but their inherent isotropic interactions prevent spontaneous elongation \cite{Kartashov2018}. This isotropic interaction is partly due to the dual-component stabilization mechanism, which utilizes magnetic Feshbach resonance (MFR) with dc magnetic field which
alters the \textit{s}-wave scattering length $a_{s}$ of the inter-atomic
interaction to achieve uniform modulation of the interaction \cite{Tiesinga1993,Moerdijk1995,Courteille1998,Bertozzi1998,Inouye1998,Timmermans1999,Duine2004,kohler2006,Jones2006,Randy,Chin2010}. In stark contrast, the characteristic feature of optical Feshbach resonance (OFR) - its ability to spatially inhomogeneously modulate scattering lengths \cite{Fisher1989,Fedichev1996,Caputo,Saito2003,Qi2011,Chien2012} - reveals an extraordinary possibility for engineering elongated vortex states. It is worth noting that recent advancements in optical manipulation demonstrate OFR's capacity for micro scale spatial density modulation in quantum droplets \cite{Schafer,Fu}, enabling precise morphology design, tailored macroscopic mechanical properties, and novel approaches for droplet dynamics manipulation. This technological progress establishes OFR as a transformative tool for quantum droplet engineering, particularly in realizing elongated vortex configurations.

The objective of the present work is to predict
\emph{stable} vortex QDs with a \textquotedblleft slender" (elongated) shape
in the binary condensate with the contact interactions, using an effective
\textit{nonlinear potential} \cite{RMP} (alias \textit{pseudopotential
}\cite{pseudo}), that may be induced by a spatially modulated interaction
strength. In turn, the modulation can be imposed by means of the spatially
inhomogeneous Feshbach resonance - OFR, it can be readily induced by
spatially non-uniform optical fields with inhomogeneous intensity $I\left(
x,y,z\right) $, which can be used to shape effective nonlinear potentials
through the local OFR relation,
\begin{equation}
a_{\mathrm{opt}}\varpropto -I\left( x,y,z\right) ,  \label{I}
\end{equation}%
where $a_{\mathrm{opt}}$ is the OFR-induced change of the scattering length,
$\left( x,y,z\right) $ are spatial coordinates, and sign minus implies that
the resonant laser illumination induces the inter-atomic attraction force,
which can be achieved by adjusting the detuning of the laser \cite%
{Theis2004,Bauer2009,Yamazaki2010,RMP,Yan2013}.

\section{Model}
In this work, we consider the possibility to use OFR for mapping the laser
intensity distribution in the usual Gaussian laser beam, propagating along
the $z$ axis, into an effective nonlinear trapping potential with the
transverse Gaussian profile, determined as per Eq. (\ref{I}), \textit{viz}.,%
\begin{equation}
a_{\mathrm{opt}}(x,y)=-a^{(0)}\exp \left(-\rho^{2}/L^{2}\right), \
\rho\equiv \sqrt{x^{2}+y^{2}},  \label{aopt}
\end{equation}
where $L$ is the beam waist, and $-a^{(0)}<0$ is the
amplitude of the OFR-induced modulation of the local scattering length. The
objective is to use the axisymmetric profile (\ref{aopt}) to stabilize
vortex solitons with the angular momentum directed along the $z$ axis and a
\textquotedblleft slender" (elongated) shape. The setting will be also used
to study more sophisticated dynamical regimes, such as gyroscopic motion
(precession) of the vortex soliton.

Our study, referring to the use of $^{39}$K atoms, implies large spectral
detuning between the OFR wavelength and that corresponding to the atomic
dipole transition, which makes it possible to neglect a linear potential
induced by the OFR field \cite{Banerjee,Campbell}. In this case, $\psi _{1,2}
$ are the wave functions of the two species, the total energy of the binary
BEC being:
\begin{gather}
E=\int_{V}\left[ \sum_{j}\left( \frac{\hbar ^{2}|\nabla \psi _{j}|^{2}}{2m}+%
\frac{\mathcal{G}_{jj}}{2}|\psi _{j}|^{4}\right) +\mathcal{G}_{12}|\psi
_{1}|^{2}|\psi _{2}|^{2}\right.   \notag \\
\phantom{=\;\;}\left. +\frac{8m^{3/2}}{15\pi ^{2}\hbar ^{3}}\left( \sum_{j}{%
\mathcal{G}_{jj}}|\psi _{j}|^{2}\right) ^{5/2}\right] \ ,\ \ j=1,2.
\label{energy}
\end{gather}%
Here $\hbar $ and $m$ are the Planck's constant and atomic mass, while $%
\mathcal{G}_{jj}=4\pi \hbar ^{2}a_{jj}/{m}$ ($j=1,2$) and $\mathcal{G}%
_{12}=4\pi \hbar ^{2}a_{12}/{m}$ are the strengths of the intra- and
inter-component mean-field interactions, respectively. Here, $a_{11},a_{22}$
and $a_{12}$ are the corresponding $s$-wave scattering lengths of atomic
collisions, and the last term of Eq. (\ref{energy}) accounts for the LHY
correction \cite{Petrov2015,Fischer2006,Inguscio2018}.

For this system, the minimization of mean-field energy is attained at $%
n_{1}/n_{2}=\sqrt{a_{22}/a_{11}}$ (where $n_{1,2}=|\psi_{1,2}|^{2}$) \cite{Petrov2015,Inguscio2018}. Here, we
consider the symmetric case with $a_{22}=a_{11}$, which leads to the
condition of $n_{2}=n_{1}$, hence the two wave-function coalesce into a
single one as $\psi _{1}=\psi _{2}\equiv \psi /\sqrt{2}$. We consider the
setup with the inter-component scattering length engineered by means of OFR
so as to set $a_{12}=a_{\mathrm{bg}}+a_{\mathrm{opt}}$, where $a_{\mathrm{bg}}<0$ is the background value of the scattering length in the absence of the
laser illumination. It is adjusted by the additional magnetic Feshbach
resonance as $a_{\mathrm{bg}}=-a_{11}$, to cancel the spatially uniform
scattering length (the detailed consideration of the two-component asymmetry
is presented in the Appendix A). Thus, energy (\ref{energy}) is
simplified to
\begin{equation}
E=\int_{V}\left( \frac{\hbar ^{2}}{2m}|\nabla \psi |^{2}+\frac{\pi \hbar
^{2}a_{\mathrm{opt}}}{m}|\psi |^{4}+\frac{2}{5}\Gamma |\psi |^{5}\right) ,
\end{equation}%
where $\Gamma =4m^{3/2}\mathcal{G}_{11}^{5/2}/3\pi ^{2}\hbar ^{3}$. Accordingly,
the dynamics is governed by the single Gross-Pitaevskii (GP) equation:
\begin{equation}
i\hbar \frac{\partial \psi }{\partial t}=-\frac{\hbar ^{2}\nabla ^{2}}{2m}%
\psi -\frac{2\pi \hbar ^{2}a^{(0)}}{m}e^{-\frac{\rho ^{2}}{L^{2}}}|\psi
|^{2}\psi +\Gamma |\psi |^{3}\psi .  \label{psi}
\end{equation}%
Stationary solutions of Eq. (\ref{psi}) with chemical potential $\mu $ and
integer vorticity $S$ are looked for in the usual form,
\begin{equation}
\psi \left( x,y,z,t\right) =\exp \left( -i\mu t+iS\theta \right) \phi \left(
\rho ,z\right) ,  \label{S}
\end{equation}%
with angular coordinate $\theta $ and positive real
function $\phi \left( \rho ,z\right) $ satisfying the stationary equation

\begin{eqnarray}
\hbar \mu \phi =-\frac{\hbar ^{2}}{2m}\nabla^{2}_{\rho z} \phi-\frac{2\pi
\hbar ^{2}a^{(0)}}{m}e^{-\frac{\rho^2}{L^{2}}} \phi ^{3}+\Gamma \phi ^{4}.
\label{phi}
\end{eqnarray}%
Where $\nabla^{2}_{\rho
z}=\partial^{2}_{\rho}+\rho^{-1}\partial_{\rho}+\partial^{2}_{z}-S^{2}/%
\rho^{2}$, Stationary solutions are characterized by the total number of
atoms, $N\equiv 2\pi \int_{0}^{\infty }\rho d\rho \int_{-\infty }^{+\infty
}dz\phi ^{2}\left( \rho,z\right) $. In the following discussions, we select $a_{11}=50a_{0}$ ($a_{0}$ is the Bohr radius) for simulations. The
lowest-energy solutions with given vorticity $S$ obtained via imaginary-time
method in the cylindrical coordinates coincide with those in the Cartesian
coordinates \cite{Mottelson1999,Bertsch1999,Wu2001}. Their stability was tested by real-time simulations
of the perturbed evolution, perforemed for a sufficient long time, which exceeds 100 ms
(much longer than the levitation time in the experiment \cite{Inguscio2018}).

\section{Characteristics of the Elongated states}
\begin{figure}[tbph]
{\includegraphics[width=1\columnwidth]{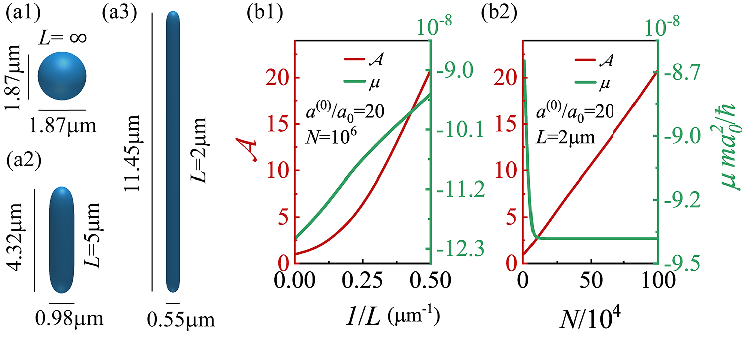}}
\caption{(a1-a3) Typical examples of fundamental ($S=0$) QDs, composed of $%
(a^{(0)}, N)=(20a_{0},10^{6})$ atoms and shaped by the OFR-induced spatial
modulation (\protect\ref{aopt}) with $L=\infty$ (i.e., in the free space)
(a1), $5\ \upmu\mathrm{m}$ (a2) and $2\ \upmu\mathrm{m}$ (a3). The vertical
and horizontal lengths, $D_{v}$ and $D_{h}$, of the QDs are indicated in the
panels. (b1) and (b2): The aspect ratio $\mathcal{A}$ and chemical potential
$\protect\mu $ of the QDs vs. $L$ and $N$, respectively. }
\label{fundamentalExample}
\end{figure}

To quantify the slenderness of 3D vortex QDs produced by
Eq. (\ref{psi}), we define the aspect ratio, $\mathcal{A}=D_{\mathrm{v}}/D_{%
\mathrm{h}}$, where $D_{\mathrm{v}}$ and $D_{\mathrm{h}}$ are, respectively,
vertical and horizontal lengths of the QDs. First, the numerical solution of
the stationary version of Eq. (\ref{psi}) for fundamental QDs [with $S=0$ in
Eq. (\ref{S})] demonstrates, in Figs. \ref{fundamentalExample}(a1-a3), the
effect of the nonlinear trapping potential on the QD's shape. In Fig. \ref%
{fundamentalExample}(a1), the aspect ratio is $\mathcal{A}=1$ for the
isotropic fundamental QD in the free space ($L=\infty $), and $%
(a^{(0)},N)=(20a_{0}, 10^{6})$. As shown in
Figs. \ref{fundamentalExample}(b1,b2), the aspect ratio increases (i.e., the
fundamental QD elongates) with the decrease of radius $L $ of the holding
beam in Eq. (\ref{aopt}) and increase of $N$. The numerical solution of the
time-dependent equation (\ref{psi}) (not shown here in detail) demonstrates
that the entire family of the elongated QDs with $S=0$ is stable. Note, in
particular, that the $\mu (N)$ dependence plotted in Fig. \ref%
{fundamentalExample}(b2) satisfies the well-known necessary stability
condition, $d\mu /dN<0$ (the Vakhitov-Kolokolov criterion \cite%
{Vakhitov1973,Berge}).

\begin{figure}[tbph]
{\includegraphics[width=1\columnwidth]{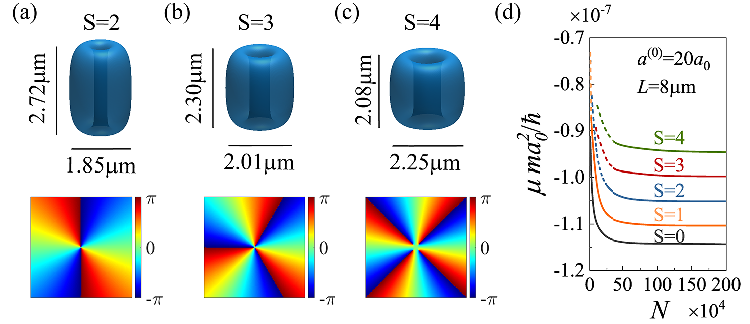}}
\caption{(a-c) Typical examples of the vortex droplets and their phases
diagrams (at $z=0$) with $(L,a^{(0)},N)=(8\ \upmu\mathrm{m}, 20a_{0}, 2\times10^{6})
$. (d) The chemical potential $\protect\mu $ of the QDs with $S=0\sim 4$ vs.
$N$, with solid and dashed lines representing stable and unstable solutions,
respectively.}
\label{S2}
\end{figure}

In Fig. \ref{fundamentalExample}(b2), the chemical potential of the
fundamental QDs initially drops steeply with the increase of $N$, and then,
at $N\rightarrow \infty $, it saturates at an equilibrium value $\mu _{0}$.
To qualitatively explain the saturation, one may take the 1D version of Eq. (%
\ref{phi}) for QDs strongly elongated in the direction of $z$, in which the
transverse $\rho $-derivatives are dropped and $S=0$ is set:%
\begin{equation}
\hbar \mu \tilde{\phi}=-\frac{\hbar ^{2}}{2m}\frac{d^{2}\tilde{\phi}}{dz^{2}}%
-\frac{2\pi \hbar ^{2}a^{(0)}}{m}\tilde{\phi}^{3}+\Gamma \tilde{\phi}^{4}.
\label{1D}
\end{equation}%
This limit corresponds to reducing Eq. (\ref{phi}) for the strongly
elongated QDs to the effectively 1D equation. It may be adopted as an
approximation corresponding to the \textquotedblleft quasi-inflexion point"
\cite{quasi-inflexion}. Further, a strongly elongated 1D solution to Eq. (%
\ref{1D}) seems as a long segment, $|z|<Z$, filled by the condensate with
the nearly constant equilibrium density $\tilde{\phi}_{0}^{2}$, which is
smoothly connected to the asymptotically zero solution, at $|z|>Z$, by kink
solutions (alias fronts) around $z=\pm Z$ \cite{Birnbaum,Petrov3}. To find
the value of $\tilde{\phi}_{0}^{2}$ and the corresponding value $\tilde{\mu}%
_{0}$ of the chemical potential, one multiplies Eq. (\ref{1D}) by $d\tilde{%
\phi}/dz$ and performs the integration, reducing the second-order equation
to the first-order one:%
\begin{equation}
\frac{1}{2}\hbar \mu \tilde{\phi}^{2}=-\frac{\hbar ^{2}}{4m}\left( \frac{d%
\tilde{\phi}}{dz}\right) ^{2}-\frac{\pi \hbar ^{2}a^{(0)}}{2m}\tilde{\phi}%
^{4}+\frac{1}{5}\Gamma \tilde{\phi} ^{5}.  \label{integrated}
\end{equation}%
The kink solution to both equations (\ref{1D}) and (\ref{integrated}),
centered at $z=Z$, satisfy the boundary conditions $\phi ^{2}(z-Z\rightarrow
-\infty )=\tilde{\phi}_{0}^{2}$ and $\tilde{\phi}^{2}(z-Z\rightarrow +\infty
)=0$. Because $z$-derivatives vanish at $z-Z\rightarrow \pm \infty $, in
this limit Eqs. (\ref{1D}) and (\ref{integrated}) amount to a system of
algebraic equations for $\tilde{\phi}_{0}^{2}$ and $\tilde{\mu}_{0}$, which
readily yields:%
\begin{equation}
\tilde{\phi}_{0}=\frac{5\hbar ^{2}a^{(0)}}{3m\Gamma },~\tilde{\mu}_{0}=-%
\frac{25\hbar ^{5}}{\Gamma ^{2}}\left( \frac{\pi a^{(0)}}{3m}\right) ^{3}.
\label{0}
\end{equation}%
The comparison of the actual saturation value $\mu _{0}$ in Fig. \ref%
{fundamentalExample}(b2) with the one approximately predicted by Eq. (\ref{0}%
) yields $\mu _{0}\approx 0.73\tilde{\mu}_{0}$, the discrepancy being
explained by the omission of the transverse structure in the above
consideration.
\begin{figure}[tbph]
{\includegraphics[width=1\columnwidth]{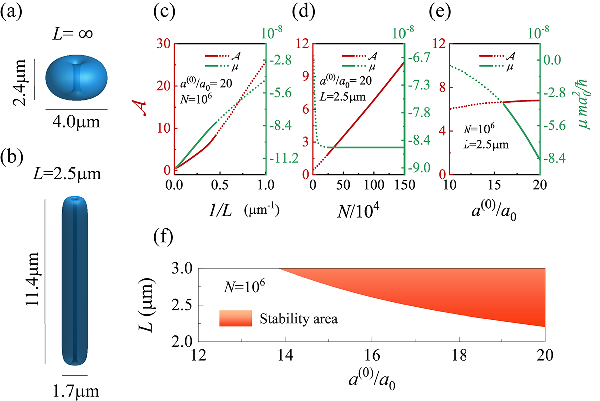}}
\caption{Characteristics of elongated vortex QDs with $S=1$. (a,b) Stable
vortex QDs in the free space ($L=\infty $) (a) and shaped by the OFR-induced
modulation with $L=2.5 \ \upmu\mathrm{m}$ (b). (c-e) The aspect ratio $%
\mathcal{A}$ and chemical potential $\protect\mu $ of the vortex QD, are
plotted by the red and green curves (which correspond, respectively, to the
left and right vertical axes) as functions $L$ (c), $N$ (d), $a^{(0)}$ (e),
respectively. In these plots, solid and dashed lines represent stable and
unstable solutions, respectively. (f) The stability chart for the vortex QDs
with $S=1$ in the plane of ($a^{(0)},L$) with $N=10^{6}$.}
\label{VortexExample}
\end{figure}
Note that the elongated QDs with embedded integer vorticity $S$ also feature
the saturation of $\mu $ at $N\rightarrow \infty $, see Fig. \ref{S2}(d)
below. As concerns the transverse structure of the strongly elongated QDs,
it obviously corresponds to solutions of the 2D reduction of Eq. (\ref{phi}%
), produced by dropping the $z$ derivative in the equation.

Next, characteristics of elongated QDs with the embedded vorticity, $S=1$,
are displayed in Fig. \ref{VortexExample}, illustrating the significant
impact of varying parameters on their shape. The solution for the 3D vortex
QD in the free space (it corresponds to $L=\infty $), which in plotted in
Fig. \ref{VortexExample}(a), coincides with the one reported in Ref. \cite%
{Kartashov2018}. Figs. \ref{VortexExample}(b) show the shapes of the
elongated vortex QDs when transverse confinement is applied. However, the
vortex QDs are unstable when $N$ is insufficiently large, or the effective
nonlinear potential is too weak, or the confinement is too tight (i.e., $%
a^{(0)}$ or $L$ is too small). In Fig. \ref{VortexExample}(c-e), the
``slenderness" and stability of the vortex QDs with $S=1$ are systematically
investigated with effects from $L$, $N$, and $a^{(0)}$. In Fig. \ref%
{VortexExample}(c), the decrease of $L$ significantly increases the aspect
ratio, making the QD unstable when $L$ falls below a critical value. Fig. %
\ref{VortexExample}(d) shows that, with the increase of $N$, the QD becomes
more elongated, with larger values of the aspect ratio. On the other hand,
below a critical value of $N$, the vortex QDs are unstable. In Fig. \ref%
{VortexExample}(e), the variation of the modulation amplitude $a^{(0)}$ has
minimal impact on the aspect ratio. The instability of the vortex QD occurs
at $a^{(0)}$ falling below a critical value. The existence of the critical
values implies that there is a stability boundary in the parametric space.
Fig. \ref{VortexExample}(f) displays the stability area in the parameter
plane of $(L, a^{(0)})$, which is produced by systematic simulations of the
perturbed evolution of the vortex QD.

\section{Dynamics}

\begin{figure}[tbph]
{\includegraphics[width=1\columnwidth]{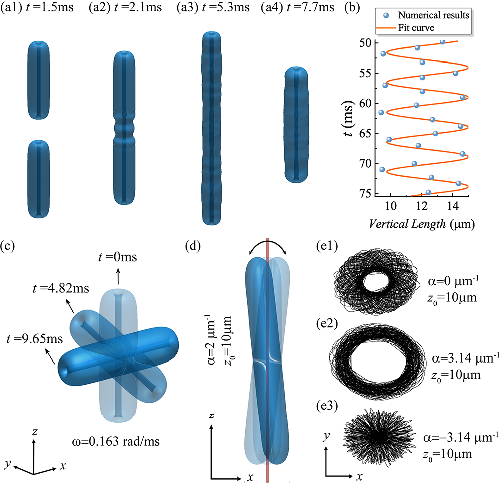}}
\caption{For all the droplets, $(L,a^{(0)}, N)=(2.5\ \upmu \mathrm{m},
20a_{0}, 5\times 10^{5})$. (a1-a4) Collision of identical elongated vortex
QDs, with $S=1$, set in motion along the $z$ axis by kicks $\protect\eta =1\
\upmu \mathrm{m^{-1}}$, the initial half-separation between the QDs being $%
z_{d}$ $=12.8\ \upmu \mathrm{m}$. Panel (b) illustrates quasi-harmonic
oscillations of the breather produced by the merger of the colliding vortex
QDs fitted to the sine function. (c) The rotation of an elongated vortex QD
driven by the rotating trapping profile (\protect\ref{omega}), with the
resulting configurations displayed at different moments, which correspond to
the rotation angles $0$, $\protect\pi /4$, and $\protect\pi /2$,
respectively. (d) The swing motion of the fundamental elongated QD in the $%
(xz)$-plane. (e1-e3) The gyroscopic motion (precession) of an elongated
vortex QD, shown by projection of the center-of-mass trajectory onto the
horizontal plane fixed at $z=2.5\ \upmu \mathrm{m}$. Panels (e1), (e2), and
(e3) display the trajectories initiated by the torque with different
strengths.}
\label{collision}
\end{figure}

The independence of the modulation profile (\ref{aopt}) on $z$ makes it
possible to generate QDs freely moving along the $z$ axis, and thus simulate
Eq. (\ref{psi}) for collisions between the QDs moving in opposite
directions, initiated by the input, $\psi _{0}\left( \rho,z\right) =\psi
(\rho,z-z_{d})\mathrm{e}^{-i\eta z}+\psi (\rho,z+z_{d})\mathrm{e}^{i\eta z}$%
, where $2z_{d}$ is the initial separation between the QDs, and $\pm \eta $
are the opposite kicks which set the droplets in motion. Figures \ref%
{collision}(a1-a4) show that the collision between the QDs with identical
vorticities $S=1$ is inelastic, resulting in the merger of the vortex QDs
into a single oscillatory one (breather). The breathing dynamics is robust
and quasi-harmonic, lasting, at least, for $>100\ \mathrm{ms}$, as is
illustrated in Fig. \ref{collision}(b) a sequence of $5$ cycles within the
time interval of $50\thicksim 75\ \mathrm{ms}$. The breathing frequency is
nearly independent of the collision velocity, suggesting that it is an
eigenfrequency of a QD's intrinsic mode. The collision leads to full
destruction of the vortex QDs if the initial kick exceeds a certain critical
value, which is $\eta_{\mathrm{cr}}=1.2\ \upmu\mathrm{m^{-1}}$. Collisions
between zero-vorticity QDs produce similar results (not shown here in
detail). On the other hand, for the QDs with opposite vorticities, $S=\pm 1$%
, the simulations demonstrate that the collision is quasi-elastic, leading
to temporary fusion into a single droplet, which then splits back into
original moving vortex QDs. The latter result is explained by the fact that
the opposite vorticities attenuate the coherent interaction between the
colliding objects. It is shown in Fig. \ref{collision}(c) that the holding
beam rotating around the $y$-axis in the $\left( xz\right) $-plane at
angular velocity $\omega =0.163\ \mathrm{rad/ms}$, which means the
replacement of the trapping profile (\ref{aopt}) by
\begin{equation}
a_{\mathrm{opt}}=-a^{(0)}\exp [-\frac{(x\cos (\omega t)-z\sin (\omega
t))^{2}+y^{2}}{L^{2}}],  \label{omega}
\end{equation}
is able to drive the adiabatic rotation of the elongated vortex QD. When the
angular velocity $\omega$ exceeds the threshold value $0.977\ \mathrm{rad/ms}
$, the enforced rotation causes the vortex droplet to split.

In addition, it is possible to excite swinging motion of an elongated QD in
the stationary trapping profile (\ref{aopt}) by applying an original torque
to it. First, Fig. \ref{collision}(d) shows that the torque, in the form of
the multiplication of $\psi \left( x,y,z\right) $ by $\exp [i\alpha x\tanh
\left( {z/z_{0}}\right) ]$ with strength $\alpha $, applied to a fundamental
(zero-vorticity) elongated QD, sets it in the state of swinging motion in
the $(xz)$-plane, with a period weakly depending on $\alpha $. In the case
displayed in Fig. \ref{collision}(d) the period is $0.792\ \mathrm{ms}$, for
$(\alpha, z_0)=(2\ \upmu \mathrm{m^{-1}}, 10\ \upmu\mathrm{m})$.
Furthermore, the application of the torque to an elongated vortex QD\
initiates its gyroscopic motion (precession), cf. Ref. \cite{gyro}. In the
present case, this is done by suddenly tilting the vortex QD in the $(xz)$%
-plane. Then, it is either released or is additionally subjected to the
action of the torque represented by factor $\exp [i\alpha y\tanh (z/z_{0})]$%
. Because the original vorticity is counter-clockwise with respect to the
positive $z$-axis, this torque with $\alpha >0$ or $\alpha <0$ makes the
vortex angular momentum smaller or larger, respectively. The precession
initiated by this procedure is represented in Figs. \ref{collision}(e1-e3)
by the projection of trajectories of the QD's center-of-mass onto the
horizontal plane fixed at $z=2.5\ \upmu \mathrm{m}$, in the time interval $%
t=9\thicksim 90\ \mathrm{ms}$. In the case of zero torque ($\alpha =0$),
Fig. \ref{collision}(e1) exhibits an elliptic trajectory rotating around the
center. For $\alpha >0$ and $\alpha <0$, Figs. \ref{collision}(e2) and (e3)
show that the trajectories become, respectively, more circular or more
elliptical when the applied torque makes the resulting angular momentum
smaller or larger, resulting the gyroscopic motion may be considered as a an
example of macroscopic spin-orbit coupling.

\section{Higher-order vortex states}
The system modeled by Eq. (\ref{psi}) also supports\emph{\ stable} QDs with
higher values of the embedded vorticity (topological charge), $S=2$, $3$,
and $4$, examples of which are displayed in Figs. \ref{S2}(a-c). It is
relevant to stress that this funding is an advantage of the nonlinear
trapping potential, as the linear 3D harmonic-oscillator one can stabilize
solely vortex solitons with $S=1$ \cite{Mihalache}. Naturally, the increase
of $S$ for fixed parameters $L$ and $a^{(0)}$ in the modulation profile (\ref%
{aopt}), and fixed number of atoms $N$ makes the radius of the vortex' inner
hole larger, leading to a decrease in the aspect ratio $\mathcal{A}$.
Families of the higher-order vortex QDs are characterized in Fig. \ref{S2}%
(d) by respective dependences of the chemical potential $\mu $ on $N$, cf.
similar dependences for $S=0$ and $1$ displayed in Figs. \ref%
{fundamentalExample}(b2) and \ref{VortexExample}(d), respectively.
Naturally, Fig. \ref{S2}(d) demonstrate that higher topological charges
produce larger higher chemical potentials. This conclusion can be easily
explained if one adopts the simple ansatz for the transverse shape of the
vortex QD, $\phi \sim \rho ^{S}\exp \left( -\rho ^{2}/\rho _{0}^{2}\right) $%
. Indeed, substituting it in Eq. (\ref{phi}) one readily obtains the $S$%
-dependent shift of the chemical potential, $\delta \mu =\left( 2\hbar
/m\rho _{0}^{2}\right) S$. The respective splitting between the eigenvalues
corresponding to vorticities $S$ and $S+1$, $\Delta \mu =2\hbar /m\rho
_{0}^{2}$, very well matches the numerical results displayed in Fig. \ref{S2}%
(d).

\section{Conclusion}
We have proposed to use the OFR, modulating the
inter-component atomic $s$-wave scattering length in the binary BEC, to
induce the effective 2D nonlinear trapping potential with the Gaussian
profile. It supports elongated QDs, including ones with embedded integer
vorticity $S$, which are stabilized by the GP-LHY equations. We have
systematically explored the effects of the modulation amplitude and width,
as well as the total number of atoms, on the shape and stability of the QDs.
The fundamental QDs with $S=0$ are completely stable, while families of the
vortex QDs with $S=1,2,3,$ and $4$ are partly stable. Collisions of
identical elongated vortex QDs with $S=1$ lead to their merger into
breathers with the same vorticity, $S=1$, while collisions between the
elongated QDs with opposite vorticities, $S=\pm 1$, are quasi-elastic. The
moving OFR modulation profile can rotate the elongated QDs in the abiabatic
regime. The application of the torque to the trapped elongated vortex QD
sets it in a state of stable gyroscopic motion (precession).

As an extension of the analysis, it may be interesting to consider more
complex holding potentials, such as anisotropic ones, as well as those
combining the nonlinear trap and vortex laser beams. Such
modes with long inner voids can be used, in the context of the currently
developing field matter-wave atomtronics, as conduits (waveguides) for
linear atomic jets, similar to the concept of "light guided by light", which
is wellknown in nonlinear optics \cite{Demircan,Vasa,Fang}.
Magnetic-gradient/radiofrequency-mediated scattering-length control,
combined with droplet-creation schemes, could simplify future experiments
\cite{Carli,Sanz,Lavoine}.

\begin{acknowledgments}
We appreciate valuable discussions with Jingxuan Sun. This work was
supported by NNSFC (China) through Grants No. 12274077, No. 12475014,
Guangdong Basic and Applied Basic Research Foundation 2023A1515110198,
2024A1515030131, 2025A1515011128, No. 2023A1515010770, the Research Fund of
Guangdong-Hong Kong-Macao Joint Laboratory for Intelligent Micro-Nano
Optoelectronic Technology through grant No.2020B1212030010. The work of
B.A.M. was supported, in part, by the Israel Science Foundation through
grant No. 1695/2022.
\end{acknowledgments}

\appendix
\section{Two-component asymmetry situation}
According to the defination of the total energy (see Eq. (3) in the main text), the corresponding two-component GP-LHY equation is written as:
\begin{align}
i\hbar \frac{\partial \psi _1}{\partial t}=-\frac{\hbar ^2}{2m}\nabla ^2\psi _1+\left( \mathcal{G}_{11}|\psi _1|^2+\mathcal{G}_{12}|\psi _2|^2 \right) \psi _1 \notag\\
+\gamma a_{11}\left( a_{11}|\psi _1|^2+a_{22}|\psi _2|^2 \right) ^{3/2}\psi _1,
\\
i\hbar \frac{\partial \psi _2}{\partial t}=-\frac{\hbar ^2}{2m}\nabla ^2\psi _2+\left( \mathcal{G}_{22}|\psi _2|^2+\mathcal{G}_{12}|\psi _1|^2 \right) \psi _2 \notag\\
+\gamma a_{22}\left( a_{11}|\psi _1|^2+a_{22}|\psi _2|^2 \right) ^{3/2}\psi _2,
\label{GP}
\end{align}

where $\gamma =128\sqrt{\pi}\hbar ^2/3m$, and we consider the setup with the inter-component scattering length engineered by means of OFR as $a_{12}=a_{\mathrm{bg}}+a_{\mathrm{opt}}=a_{\mathrm{bg}}-a^{\left( 0 \right)}\exp\left(-\rho ^2/L^2\right)$ (where $\rho \equiv \sqrt{x^2+y^2}$). Here, $a_{\mathrm{bg}}<0$ is the background value of the scattering length in the absence of light, which is adjusted by the additional magnetic Feshbach resonance and characterizes an attractive interaction between the two species of $\psi_1$ and $\psi_2$. Further, $L$ is the beam waist, and $-a^{\left( 0 \right)}<0$ is the amplitude of the OFR-induced modulation of the local scattering length, the total numbers of atoms in the two components are
\begin{align}
N_1\equiv \int{|\psi _1|^2dxdydz},\,\,N_2\equiv \int{|\psi _2|^2dxdydz}.
\label{N}
\end{align}

Regarding the asymmetric situation discussed here, under the requirement of the minimization of the mean-field energy, the ratio of the equilibrium densities is locked as $n_1/n_2=\sqrt{\mathcal{G}_{22}/\mathcal{G}_{11}}$ \cite{Petrov2015}, hence the ratio of the particle numbers is also $N_1/N_2=\sqrt{a_{22}/a_{11}}$ \cite{Inguscio2018}, which results in $a_{22}=\left( N_1/N_2 \right) ^2a_{11}$. Here we keep fixing $a_{\mathrm{bg}}=-a_{11}$ and simulate the elongated vortex with the imbalanced density by adjusting the total number of atoms $N_{1,2}$ in each of the two components.

\begin{figure}[htbp]
{\includegraphics[width=0.6\columnwidth]{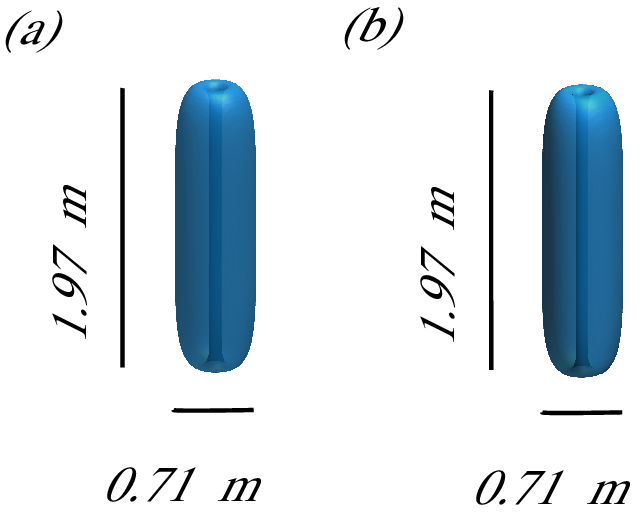}}
\caption{Typical examples of the asymmetric population imbalance for the two-components system, with $N_{1}=2.4\times10^{5}$, $N_{2}=2.6\times10^{5}$, $L=2.5\mathrm{\mu m}$, $a^{(0)}=20a_{0}$, and $a_{11}=50a_{0}$. (a) and (b) represent the component $\psi_{1}$ and $\psi_{2}$, respectively.}\label{F1}
\end{figure}

First, we consider the situation with slight imbalances. As shown in Fig. (\ref{F1}), we set $N_{1}=2.4\times10^{5},N_{2}=2.6\times10^{5}$, we have constructed the corresponding imbalanced slender vortex solution and demonstrated its stability through direct evolution.

\begin{figure}[htbp]
{\includegraphics[width=0.85\columnwidth]{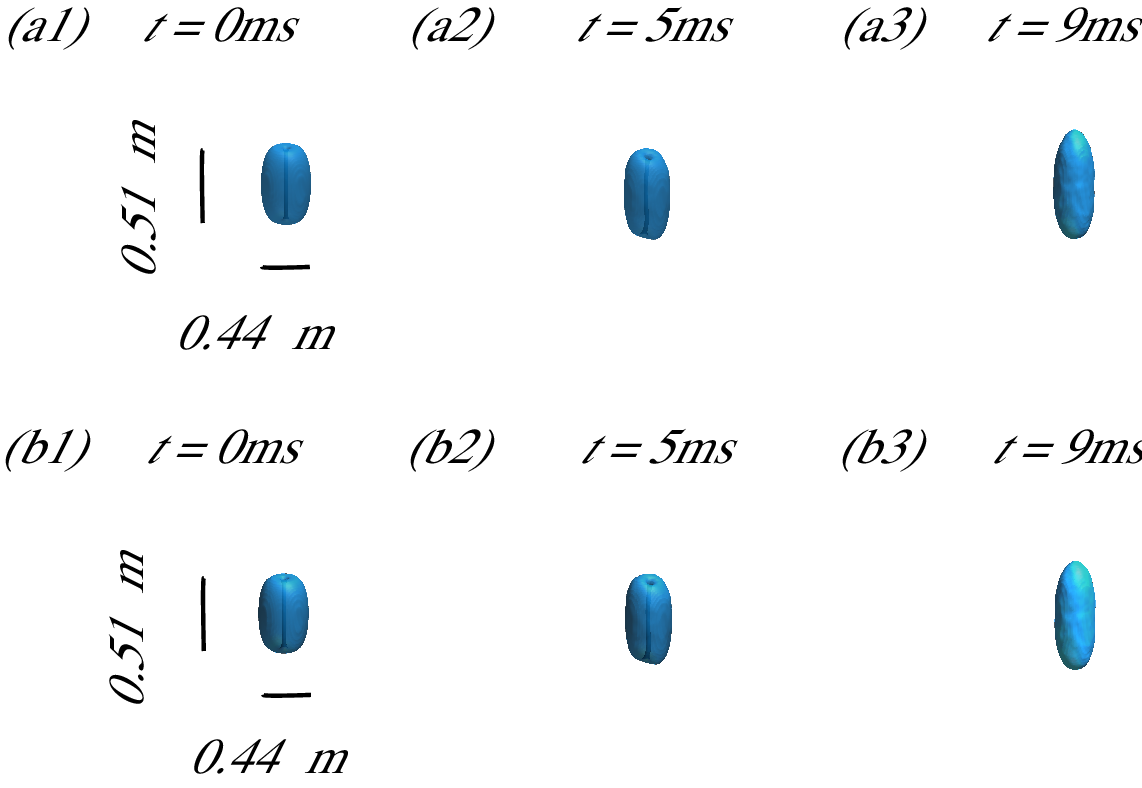}}
\caption{Typical examples and evolution of the unstable solution, with $N_{1}=2\times10^{5}$, $N_{2}=3\times10^{5}$, $L=2.5\mathrm{\mu m}$, $a^{(0)}=20a_{0}$, and $a_{11}=50a_{0}$. (a1-a3) and (b1-b3) represent the component $\psi_{1}$ and $\psi_{2}$, respectively.
}\label{F2}
\end{figure}

Then, we consider the situation with slight imbalances. We set $N_{1}=2\times10^{5},N_{2}=3\times10^{5}$, the corresponding imbalanced slender vortex solution is unstable, the direct evolution as shown in Fig. (\ref{F2})

\FloatBarrier

The above results show that the size profiles of the two components are the same, but the densities are different. Overall, we have found that the stable vortices persist if the population imbalance between the two components is not too large. On the other hand, if the intercomponent population imbalance attains a critical value, the vortex QD becomes unstable and decays to the ground state within a few microseconds.


\begin{thebibliography}{99}
\bibitem{McWilliams1994a} J. C. McWilliams, J. B. Weiss, and I. Yavneh,
Anisotropy and Coherent Vortex Structures in Planetary Turbulence, Science
\textbf{264}, 410 (1994). 

\bibitem{astro} J. A. Barranco and P. S. Marcus, Three-dimensional vortices
in stratified protoplanetary disks, Astrophys. Journal \textbf{623},
1157-1170 (2005).

\bibitem{McWilliams1994b} J. C. McWilliams and J. B. Weiss, Anisotropic
geophysical vortices, Chaos: An Interdisciplinary Journal of Nonlinear
Science \textbf{4}, 305 (1994). 


\bibitem{plasma1} M. Tajiri and H. Maesono, Resonant interactions of drift
vortex solitons in a convective motion of a plasma, Phys. Rev. E \textbf{55}%
, 3351-3357 (1997).

\bibitem{plasma2} A. D. Rogava, G. D. Chagelishvili,and S. M. Mahajan, Shear
Langmuir vortex: An elementary mode of plasma collective behavior, Phys.
Rev. E \textbf{57}, 7103-7110 (1998).

\bibitem{plasma3} P. K. Shukla, A. A. Mamun, and B. Eliasson, 3D
electron-acoustic solitary waves introduced by phase space electron vortices
in magnetized space plasmas, Geophys. Res. Lett. \textbf{31}, L07803 (2004).

\bibitem{Dolan1989} G. J. Dolan, F. Holtzberg, C. Feild, and T. R. Dinger,
Anisotropic vortex structure in ${\mathrm{Y}}_{1}$${\mathrm{Ba}}_{2}$${%
\mathrm{Cu}}_{3}$${\mathrm{O}}_{7}$, Phys. Rev. Lett. \textbf{62} 2184
(1989).

\bibitem{Odobesko2020} A. Odobesko, F. Friedrich, S.-B. Zhang, S. Haldar, S.
Heinze, B. Trauzettel, and M. Bode, Anisotropic vortices on superconducting
Nb(110), Phys. Rev. B \textbf{102}, 174502 (2020).

\bibitem{Hung2012} H.-H. Hung, C.-L. Song, X. Chen, X. Ma, Q. Xue, and C.
Wu, Anisotropic vortex lattice structures in the FeSe superconductor, Phys.
Rev. B \textbf{85}, 104510 (2012).

\bibitem{Olszewski2018} M. W. Olszewski, M. R. Eskildsen, C. Reichhardt, and
C. J. O. Reichhardt, Structural transitions in vortex systems with
anisotropic interactions, New J. Phys. \textbf{20}, 023005 (2018).


\bibitem{Freund1997} I. Freund and V. Freilikher, Parameterization of
anisotropic vortices, J. Opt. Soc. Am. A \textbf{14}, 1902 (1997).

\bibitem{Sharma2011} M. K. Sharma, J. Joseph, and P. Senthilkumaran,
Selective edge enhancement using anisotropic vortex filter, Appl. Opt.
\textbf{50}, 5279 (2011). 


\bibitem{Stockhofe2011} J. Stockhofe, S. Middelkamp, P. G. Kevrekidis, and
P. Schmelcher, Impact of anisotropy on vortex clusters and their dynamics,
EPL \textbf{93}, 20008 (2011).

\bibitem{Liu2013} C.-F. Liu, Y.-M. Yu, S.-C. Gou, and W.-M. Liu, Vortex
chain in anisotropic spin-orbit-coupled spin-1 Bose-Einstein condensates,
Phys. Rev. A \textbf{87}, 063630 (2013).

\bibitem{McEndoo2010} S. McEndoo and Th. Busch, Vortex dynamics in
anisotropic traps, Phys. Rev. A \textbf{82}, 013628 (2010).

\bibitem{OKtel2004} M. \"{O}. Oktel, Vortex lattice of a Bose-Einstein
condensate in a rotating anisotropic trap, Phys. Rev. A \textbf{69}, 023618
(2004).

\bibitem{Engels2002} P. Engels, I. Coddington, P. C. Haljan, and E. A.
Cornell, Nonequilibrium Effects of Anisotropic Compression Applied to Vortex
Lattices in Bose-Einstein Condensates, Phys. Rev. Lett. \textbf{89}, 100403
(2002).

\bibitem{Tikhonenkov2008} I. Tikhonenkov, B. A. Malomed, and A. Vardi,
Anisotropic Solitons in Dipolar Bose-Einstein Condensates, Phys. Rev. Lett.
\textbf{100}, 090406 (2008).

\bibitem{Tan2023} Z. Tan, H. Gong, B. Zhu, H. Zhong, and S. Hu,
Two-dimensional anisotropic vortex-bright soliton and its dynamics in
dipolar Bose-Einstein condensates in optical lattice, Nonlinear Dyn. \textbf{%
111}, 9467 (2023).


\bibitem{LHY} T. D. Lee, K. Huang, and C. N. Yang, Eigenvalues and
eigenfunctions of a Bose system of hard spheres and its low-temperature
properties, Phys. Rev. \textbf{106}, 1135-1145 (1957).

\bibitem{Petrov2015} D. S. Petrov, Quantum Mechanical Stabilization of a
Collapsing Bose-Bose Mixture, Phys. Rev. Lett. \textbf{115}, 155302 (2015).

\bibitem{Petrov2016} D. S. Petrov and G. E. Astrakharchik, Ultradilute
Low-Dimensional Liquids, Phys. Rev. Lett. \textbf{117}, 100401 (2016).

\bibitem{Luca1} A. Cappellaro, T. Macr\'{\i}, G. F. Bertacco, and L.
Salasnich, Equation of state and self-bound droplet in Rabi-coupled Bose
mixtures, Sci. Rep. \textbf{7}, 13358 (2017).

\bibitem{Jrgensen2018} N. B. J\o {}rgensen, G. M. Bruun, and J. J. Arlt,
Dilute Fluid Governed by Quantum Fluctuations, Phys. Rev. Lett. \textbf{121}%
, 173403 (2018).

\bibitem{Cabrera2018} C. R. Cabrera, L. Tanzi, J. Sanz, B. Naylor, P.
Thomas, P. Cheiney, and L. Tarruell, Quantum Liquid Droplets in a Mixture of
Bose-Einstein Condensates, Science \textbf{359}, 301 (2018).

\bibitem{Ferrier-Barbut2018} I. Ferrier-Barbut and T. Pfau, Quantum Liquids
Get Thin, Science \textbf{359}, 274-275 (2018).

\bibitem{Cheiney2018} P. Cheiney, C. R. Cabrera, J. Sanz, B. Naylor, L.
Tanzi, and L. Tarruell, Bright Soliton to Quantum Droplet Transition in a
Mixture of Bose-Einstein Condensates, Phys. Rev. Lett. \textbf{120}, 135301
(2018).

\bibitem{Inguscio2018} G. Semeghini, G. Ferioli, L. Masi, C. Mazzinghi, L.
Wolswijk, F. Minardi, M. Modugno, G. Modugno, M. Inguscio, and M. Fattori,
Self-Bound Quantum Droplets of Atomic Mixtures in Free Space, Phys. Rev.
Lett. \textbf{120}, 235301 (2018).

\bibitem{Luca2} C. D'Errico, A. Burchianti, M. Prevedelli, L. Salasnich, F.
Ancilotto, M. Modugno, F. Minardi, and C. Fort, Observation of quantum
droplets in a heteronuclear bosonic mixture, Phys. Rev. Res. \textbf{1},
033155 (2019).

\bibitem{Yin2021} Y.-C. Xiong and L. Yin, Self-Bound Quantum Droplet with
Internal Stripe Structure in 1D Spin-Orbit-Coupled Bose Gas, Chin. Phys.
Lett. \textbf{38}, 070301 (2021).

\bibitem{dlw2021} L. Dong and Y. V. Kartashov, Rotating Multidimensional
Quantum Droplets, Phys. Rev. Lett. \textbf{126}, 244101 (2021).

\bibitem{Luo2021} Z. Luo, W. Pang, B. Liu, Y. Li, and B. A. Malomed, A new
kind form of liquid matter: Quantum droplets, Front. Phys. \textbf{16},
32201 (2021).

\bibitem{Cui2025} Y. Ma and X. Cui, Shell-Shaped Quantum Droplet in a
Three-Component Ultracold Bose Gas, Phys. Rev. Lett. \textbf{134}, 043402
(2025).

\bibitem{Kartashov2018} Y. V. Kartashov, B. A. Malomed, L. Tarruell, and L.
Torner, Three-dimensional droplets of swirling superfluids, Phys. Rev. A
\textbf{98}, 013612 (2018).

\bibitem{QDvortex2D} Y. Li, Z. Chen, Z. Luo, C. Huang, H. Tan, W. Pang, and
B. A. Malomed, Two-dimensional vortex quantum droplets, Phys. Rev. A \textbf{%
98}, 063602 (2018).

\bibitem{Tengstrand2019} M. N. Tengstrand, P. St\"{u}rmer, E. \"{O}.
Karabulut, and S. M. Reimann, Rotating Binary Bose-Einstein Condensates and
Vortex Clusters in Quantum Droplets, Phys. Rev. Lett. \textbf{123}, 160405
(2019).

\bibitem{Chen2024} G. Chen, H. Wang, H. Deng, and B. A. Malomed, Vortex
quantum droplets under competing nonlinearities, Chin. Phys. Lett. \textbf{41%
}, 020501 (2024).

\bibitem{Li2024FOPview} G. Li, Z. Zhao, B. Liu, Y. Li, Y. V. Kartashov, and
B. A. Malomed, Can vortex quantum droplets be realized experimentally?,
Front. Phys. \textbf{20}, 013401 (2025).

\bibitem{Stuhler2005} J. Stuhler, A. Griesmaier, T. Koch, M. Fattori, T.
Pfau, S. Giovanazzi, P. Pedri, and L. Santos, Observation of Dipole-Dipole
Interaction in a Degenerate Quantum Gas, Phys. Rev. Lett. \textbf{95},
150406 (2005).

\bibitem{Lahaye2009} T. Lahaye, C. Menotti, L. Santos, M. Lewenstein, and T.
Pfau, The Physics of Dipolar Bosonic Quantum Gases, Rep. Prog. Phys. \textbf{%
72}, 126401 (2009).

\bibitem{Ferrier-Barbut2016} I. Ferrier-Barbut, H. Kadau, M. Schmitt, M.
Wenzel, and T. Pfau, Observation of Quantum Droplets in a Strongly Dipolar
Bose Gas, Phys. Rev. Lett. \textbf{116}, 215301 (2016).

\bibitem{Schmitt2016} M. Schmitt, M. Wenzel, F. B\"{o}ttcher, I.
Ferrier-Barbut, and T. Pfau, Self-Bound Droplets of a Dilute Magnetic
Quantum Liquid, Nature \textbf{539}, 259 (2016).

\bibitem{Santos} L. Chomaz, S. Baier, D. Petter, M. J. Mark, F. Wachtler, L.
Santos, and F. Ferlaino, Quantum-fluctuation-driven crossover from a dilute
Bose-Einstein condensate to a macrodroplet in a dipolar quantum fluid, Phys.
Rev. X \textbf{6}, 041039 (2016).

\bibitem{Wang2020} Y. Wang, L. Guo, S. Yi, and T. Shi, Theory for Self-Bound
States of Dipolar Bose-Einstein Condensates, Phys. Rev. Research \textbf{2},
043074 (2020).

\bibitem{Bottcher2021} F. B\"{o}ttcher, J.-N. Schmidt, J. Hertkorn, K. S. H.
Ng, S. D. Graham, M. Guo, T. Langen, and T. Pfau, New States of Matter with
Fine-Tuned Interactions: Quantum Droplets and Dipolar Supersolids, Rep.
Prog. Phys. \textbf{84}, 012403 (2021).

\bibitem{Yin2022} F. Zhang and L. Yin, Phonon Stability of Quantum Droplets
in Dipolar Bose Gases, Chin. Phys. Lett. \textbf{39}, 060301 (2022).

\bibitem{Cidrim2018} A. Cidrim, F. E. A. dos Santos, E. A. L. Henn, and T.
Macr\'{\i}, Vortices in Self-Bound Dipolar Droplets, Phys. Rev. A \textbf{98}%
, 023618 (2018).

\bibitem{Li2024PRL} G. Li, Z. Zhao, X. Jiang, Z. Chen, B. Liu, B. A.
Malomed, and Y. Li, Strongly Anisotropic Vortices in Dipolar Quantum
Droplets, Phys. Rev. Lett. \textbf{133}, 053804 (2024).

\bibitem{Li2024FOP} G. Li, X. Jiang, B. Liu, Z. Chen, B. A. Malomed, and Y.
Li, Two-Dimensional Anisotropic Vortex Quantum Droplets in Dipolar
Bose-Einstein Condensates, Front. Phys. \textbf{19}, 22202 (2024).

\bibitem{Tiesinga1993} E. Tiesinga, B. J. Verhaar, and H. T. C. Stoof,
Threshold and resonance phenomena in ultracold ground-state collisions,
Phys. Rev. A \textbf{47}, 4114 (1993).

\bibitem{Moerdijk1995} A. J. Moerdijk, B. J. Verhaar, and A. Axelsson,
Resonances in ultracold collisions of $^{6}\mathrm{Li}$, $^{7}\mathrm{Li}$,
and $^{23}\mathrm{Na}$, Phys. Rev. A \textbf{51}, 4852 (1995).

\bibitem{Courteille1998} P. Courteille, R. S. Freeland, D. J. Heinzen, F. A.
van Abeelen, and B. J. Verhaar, Observation of a Feshbach Resonance in Cold
Atom Scattering, Phys. Rev. Lett. \textbf{81}, 69 (1998).

\bibitem{Bertozzi1998} A. L. Bertozzi, A. M\"{u}nch, X. Fanton, and A. M.
Cazabat, Contact Line Stability and \textquotedblleft Undercompressive
Shocks" in Driven Thin Film Flow, Phys. Rev. Lett. \textbf{81}, 5169 (1998).

\bibitem{Inouye1998} S. Inouye, M. R. Andrews, J. Stenger, H.-J. Miesner, D.
M. Stamper-Kurn, and W. Ketterle, Observation of Feshbach Resonances in a
Bose-Einstein Condensate, Nature \textbf{392}, 151 (1998).

\bibitem{Timmermans1999} E. Timmermans, Feshbach resonances in atomic
Bose-Einstein condensates, Phys. Rep. \textbf{315}, 199 (1999).

\bibitem{Duine2004} R. A. Duine and H. T. C. Stoof, Atom-molecule coherence
in Bose gases, Phys. Rep. \textbf{396}, 115 (2004).

\bibitem{kohler2006} T. K\"{o}hler, K. G\'{o}ral, and P. S. Julienne,
Production of cold molecules via magnetically tunable Feshbach resonances,
Rev. Mod. Phys. \textbf{78}, 1311 (2006).

\bibitem{Jones2006} K. M. Jones, E. Tiesinga, P. D. Lett, and P. S.
Julienne, Ultracold photoassociation spectroscopy: Long-range molecules and
atomic scattering, Rev. Mod. Phys. \textbf{78}, 483 (2006).

\bibitem{Randy} S. E. Pollack, D. Dries, M. Junker, Y. P. Chen, T. A.
Corcovilos, and R. G. Hulet, Extreme Tunability of Interactions in a $^7$Li
Bose-Einstein Condensate, Phys. Rev. Lett. \textbf{102}, 090402 (2009).

\bibitem{Chin2010} C. Chin, R. Grimm, P. Julienne, and E. Tiesinga, Feshbach
resonances in ultracold gases, Rev. Mod. Phys. \textbf{82}, 1225 (2010).

\bibitem{Fisher1989} M. P. A. Fisher, P. B. Weichman, G. Grinstein, and D.
S. Fisher, Boson Localization and the Superfluid-Insulator Transition, Phys.
Rev. B \textbf{40}, 546 (1989).

\bibitem{Fedichev1996} P. O. Fedichev, Yu. Kagan, G. V. Shlyapnikov, and J.
T. M. Walraven, Influence of Nearly Resonant Light on the Scattering Length
in Low-Temperature Atomic Gases, Phys. Rev. Lett. \textbf{77}, 2913 (1996).

\bibitem{Caputo} F. Kh. Abdullaev, J. G. Caputo, R. A. Kraenkel, and B. A.
Malomed, Controlling collapse in Bose-Einstein condensation by temporal
modulation of the scattering length,. Phys. Rev. A \textbf{67}, 013605
(2003).

\bibitem{Saito2003} H. Saito and M. Ueda, Dynamically Stabilized Bright
Solitons in a Two-Dimensional Bose-Einstein Condensate, Phys. Rev. Lett.
\textbf{90}, 040403 (2003).

\bibitem{Qi2011} R. Qi and H. Zhai, Bound States and Scattering Resonances
Induced by Spatially Modulated Interactions, Phys. Rev. Lett. \textbf{106},
163201 (2011).

\bibitem{Chien2012} C.-C. Chien, Spatially Varying Interactions Induced in
Ultra-Cold Atoms by Optical Feshbach Resonance, Phys. Lett. A \textbf{376},
729 (2012).

\bibitem{Schafer} F. Sch\"{a}fer, T. Fukuhara, S. Sugawa, Y. Takasu, and Y.
Takahashi, Tools for quantum simulation with ultracold atoms in optical
lattices, Nat. Rev. Phys. \textbf{2}, 411 (2020).

\bibitem{Fu} X. Zhang, Y. Hu, X. Zhang, Z. Li, Z. Chen, and S. Fu, On-Demand
Subwavelength-Scale Light Sculpting Using Nanometric Holograms, Laser \&
Photonics Rev. \textbf{17}, 2300527 (2023).

\bibitem{RMP} Y. V. Kartashov, B. A. Malomed, and L. Torner, Solitons in
nonlinear lattices, Rev. Mod. Phys. \textbf{83}, 247-306 (2011).

\bibitem{pseudo} W. A. Harrison, \textit{Pseudopotentials in the Theory of
Metals} (Benjamin, New York, 1966)

\bibitem{Theis2004} M. Theis, G. Thalhammer, K. Winkler, M. Hellwig, G.
Ruff, R. Grimm, and J. H. Denschlag, Tuning the Scattering Length with an
Optically Induced Feshbach Resonance, Phys. Rev. Lett. \textbf{93}, 123001
(2004).

\bibitem{Bauer2009} D. M. Bauer, M. Lettner, C. Vo, G. Rempe, and S. D\"urr,
Control of a Magnetic Feshbach Resonance with Laser Light, Nat. Phys.
\textbf{5}, 339 (2009).

\bibitem{Yamazaki2010} R. Yamazaki, S. Taie, S. Sugawa, and Y. Takahashi,
Submicron Spatial Modulation of an Interatomic Interaction in a
Bose-Einstein Condensate, Phys. Rev. Lett. \textbf{105}, 050405 (2010).

\bibitem{Yan2013} M. Yan, B. J. DeSalvo, B. Ramachandhran, H. Pu, and T. C.
Killian, Controlling Condensate Collapse and Expansion with an Optical
Feshbach Resonance, Phys. Rev. Lett. \textbf{110}, 123201 (2013).

\bibitem{Banerjee} A. Banerjee and V. Natarajan, Absolute-frequency
measurements of the $D_{2}$ line and fine-structure interval in $^{39}K$, Phys. Rev. A
\textbf{70}, 052505 (2004).

\bibitem{Campbell} R. L. D. Campbell, R. P. Smith, N. Tammuz, S. Beattie, S.
Moulder, and Z. Hadzibabic, Efficient Production of Large 39K Bose-Einstein
Condensates, Phys. Rev. A \textbf{82}, 063611 (2010).

\bibitem{Fischer2006} R. Sch\"{u}tzhold, M. Uhlmann, Y. Xu, and U. R.
Fischer, Mean-field Expansion in Bose-Einstein Condensates with Finite-range
Interactions, Int. J. Mod. Phys. B \textbf{20}, 3555 (2006).

\bibitem{Mottelson1999} B. Mottelson, The Yrast Spectra of Weakly Interacting Bose-Einstein Condensates, Phys. Rev. Lett. \textbf{83}, 2695 (1999).

\bibitem{Bertsch1999} G. F. Bertsch and T. Papenbrock, Yrast Line for Weakly Interacting Trapped Bosons, Phys. Rev. Lett. \textbf{83}, 5412 (1999).

\bibitem{Wu2001} Y. Wu, X. Yang, and Y. Xiao, Analytical Method for Yrast Line States in Interacting Bose-Einstein Condensates, Phys. Rev. Lett. \textbf{86}, 2200 (2001).

\bibitem{Vakhitov1973} N. G. Vakhitov and A. A. Kolokolov, Stationary
solutions of the wave equation in a medium with nonlinearity saturation,
Radiophys. Quantum Electron \textbf{16}, 783 (1973).

\bibitem{Berge} L. Berg\'{e}, Wave collapse in physics: principles and
applications to light and plasma waves, Phys. Rep. \textbf{303}, 259-370
(1998).

\bibitem{quasi-inflexion} This point is chosen by condition $\left(
\partial^{2}_{\rho}+\rho^{-1}\partial_{\rho} \right) \phi =0$ applied to Eq.
(\ref{phi}). For the solution with the Gaussian transverse profile, $%
\phi\sim \exp \left( -\rho ^{2}/\rho _{0}^{2}\right) $, this point is $%
\rho=\rho _{0}$; in that case coefficient $a^{(0)}$ in Eq. (\ref{aopt})
includes an extra factor, $1/e$.

\bibitem{Birnbaum} Z. Birnbaum and B. A. Malomed, Families of spatial
solitons in a two-channel waveguide with the cubic-quintic nonlinearity,
Physica D \textbf{237}, 3252-3262 (2008).

\bibitem{Petrov3} M. Tylutki, G. E. Astrakharchik, B. A. Malomed, and D. S.
Petrov, Collective excitations of a one-dimensional quantum droplet, Phys.
Rev. A \textbf{101}, 051601(R) (2020).


\bibitem{gyro} R. Driben, Y. V. Kartashov, B. A. Malomed, T. Meier, and L.
Torner, Soliton gyroscopes in media with spatially growing repulsive
nonlinearity, Phys. Rev. Lett. \textbf{112}, 020404 (2014).

\bibitem{Mihalache} D. Mihalache, D. Mazilu, B. A. Malomed, and F. Lederer,
Vortex stability in nearly-two-dimensional Bose-Einstein condensates with
attraction. Phys. Rev. A \textbf{73}, 043615 (2006).

\bibitem{Demircan} A. Demircan, Sh. Amiranashvili, and G. Steinmeyer,
Controlling Light by Light with an Optical Event Horizon, Phys. Rev. Lett.
\textbf{106}, 163901 (2011).

\bibitem{Vasa} P. Vasa, W. Wang, R. Pomraenke, M. Lammers, M. Maiuri, C.
Manzoni, G. Cerullo, and C. Lienau, Real-time observation of ultrafast Rabi
oscillations between excitons and plasmons in metal nanostructures with
J-aggregates, Nature Photon \textbf{7}, 128 (2013).

\bibitem{Fang} X. Fang, M. Lun Tseng, J.-Y. Ou, K. F. MacDonald, D. Ping
Tsai, and N. I. Zheludev, Ultrafast all-optical switching via coherent
modulation of metamaterial absorption, Applied Physics Letters \textbf{104},
141102 (2014).

\bibitem{Carli} A. Di Carli, G. Henderson, S. Flannigan, C. D. Colquhoun, M.
Mitchell, G.-L. Oppo, A. J. Daley, S. Kuhr, and E. Haller, Collisionally
Inhomogeneous Bose-Einstein Condensates with a Linear Interaction Gradient,
Phys. Rev. Lett. \textbf{125}, 183602 (2020).

\bibitem{Sanz} J. Sanz, A. Fr\~{A}\P lian, C. S. Chisholm, C. R. Cabrera,
and L. Tarruell, Interaction Control and Bright Solitons in Coherently
Coupled Bose-Einstein Condensates, Phys. Rev. Lett. \textbf{128}, 013201
(2022).

\bibitem{Lavoine} L. Lavoine, A. Hammond, A. Recati, D. S. Petrov, and T.
Bourdel, Beyond-Mean-Field Effects in Rabi-Coupled Two-Component
Bose-Einstein Condensate, Phys. Rev. Lett. \textbf{127}, 203402 (2021).
\end{thebibliography}
\end{document}